\documentclass[a4paper]{article}

\usepackage{INTERSPEECH2019}
\usepackage{hyperref}
\usepackage{amsmath,graphicx}
\usepackage{array,booktabs,arydshln,xcolor}

\usepackage{scrextend}

\title{End-to-end music source separation:\\is it possible in the waveform domain?}
\name{Francesc Llu\'{i}s$^*$ \qquad Jordi Pons$^*$\thanks{$^*$Contributed equally.} \qquad Xavier Serra}
\address{Music Technology Group, Universitat Pompeu Fabra, Barcelona.}
\email{name.surname@upf.edu}

\begin{document}

\maketitle
\begin{abstract}
Most of the currently successful source separation techniques use the magnitude spectrogram as input, and are therefore by default omitting part of the signal: the phase. To avoid omitting potentially useful information, we study the viability of using end-to-end models for music source separation --- which take into account all the information available in the raw audio signal, including the phase.
Although during the last decades end-to-end music source separation has been considered almost unattainable, our results confirm that waveform-based models can 
perform similarly (if not better) than
a spectrogram-based deep learning model. Namely: a Wavenet-based model we propose and Wave-U-Net can outperform DeepConvSep, a recent spectrogram-based deep learning model.  
\end{abstract}

\section{Introduction}
\label{sec:intro}

When two or more sounds co-exist, they interfere with each other resulting in a novel mixture signal where sounds are superposed (and, sometimes, masked). The source separation task tackles the inverse problem of recovering each individual sound source contribution from an observed mixture signal. 

\begin{figure*}[!t]
	\centering
	\includegraphics[width=\linewidth]{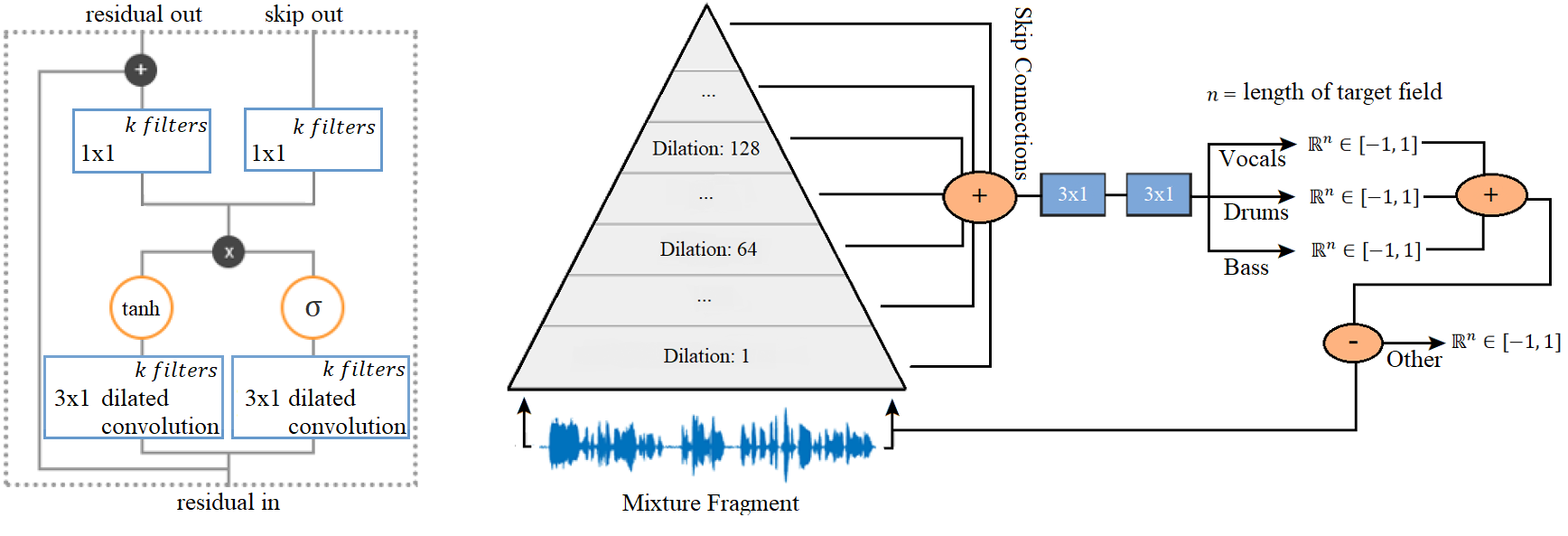}
		\vspace{-6mm}
	\caption{\textbf{Left} -- Residual layer. \textbf{Right} -- Overview of the  non-causal Wavenet we propose for multi-instrument source separation.}
	\label{fig:dos}
	\vspace{-4mm}
\end{figure*}
With the recent advances in deep learning, source separation techniques have improved substantially~\cite{stoter20182018}. 
Interestingly, though, nearly all successful deep learning algorithms use the magnitude spectrogram as input~\cite{stoter20182018,chandna2017monoaural,jansson2017singing} --- and are therefore, by default, omitting part of the signal: the phase. 
Omitting the potentially useful information of the phase entails the risk of finding a sub-optimal solution. In this work, we aim to take full advantage of the acoustic modeling capabilities of deep learning to investigate whether it is possible to approach the problem of music source separation directly in an end-to-end learning fashion. 
Consequently, our investigation is centered on studying how to separate music sources (e.g., singing voice, bass or drums) directly from the raw waveform music mixture.

During the last two decades, matrix decomposition methods have dominated the field of audio source separation.
Several algorithms have been proposed throughout the years, with independent component analysis (ICA)~\cite{hyvarinen2000independent}, sparse coding~\cite{olshausen1997sparse}, or non-negative matrix factorization (NMF)~\cite{lee2001algorithms} being the most used ones. 
Given that magnitude or power spectrogram representations are always non-negative, imposing a non-negative constraint (like in NMF) is particularly useful when analyzing these spectrograms~--- but less appropriate for processing waveforms, which range from~\mbox{-1}~to~1. For that reason, methods like ICA and sparse coding have historically been used to process waveforms \cite{dubnov2002extracting,blumensath2004unsupervised,jang2003maximum}.
Waveform representations preserve all the information available in the raw signal. However, given the unpredictable behavior of the phase in real-life sounds, it is rare to find identical waveforms produced by the same sound source. As a result of this variability, a single basis\footnote{ICA, sparse coding \& NMF model the mixture signal as a weighted sum of bases, which represent a source or components of a source.} cannot represent a sound source and therefore, one requires \textit{i)} a large amount of bases, or \textit{ii)} shift-invariant bases to obtain accurate decompositions~\cite{blumensath2004unsupervised,virtanen2006unsupervised}. 
Although several matrix decomposition methods have been used for decomposing waveform-based mixtures \cite{dubnov2002extracting,blumensath2004unsupervised,jang2003maximum}, these have never worked \mbox{as well as the spectrogram-based~ones.}

Due to the above mentioned difficulties, the phase of complex time-frequency representations is commonly discarded, assuming that magnitude spectrograms already carry meaningful information about the sound sources to be separated.
Phase related problems disappear when sounds are just represented as magnitude or power spectrograms, since different realizations of the same sound are almost identical in this time-frequency plane.
This allows to easily overcome the variability problem \mbox{found when operating with waveforms.}

Most matrix decomposition methods rely on a signal model assuming that sources add linearly in the time domain~\cite{virtanen2006unsupervised}$^1$. 

However, the addition of signals in the time and frequency domains is not equivalent if phases are discarded.
Only in expectation: $E\{|X(k)|^{2}\} = |Y_{1}(k)|^{2} + |Y_{2}(k)|^{2}$, where \(X(k) = DFT\{x(t)\}\). This means that we can approximate the time-domain summation in the power spectral domain.
For that reason, many approaches utilize power spectrograms as inputs. Although magnitude spectrograms work well in practice~\cite{roebel2015automatic}, there is no similar theoretical justification for such an inconsistency with the signal model when the phases are discarded.

Finally, note that these methods operating on top of spectrograms still need to deliver a waveform signal. To this end, the main practice is to filter the original magnitude or power spectrogram with (predicted) time-frequency masks. Accordingly, the original noisy phase of the mixture is used when synthesizing the waveform of the estimated sources --- which might introduce an additional source of error~\cite{virtanen2006unsupervised}. Notably, many modern spectrogram-based deep learning models are also relying on this same (potentially problematic) approach~\cite{chandna2017monoaural,huang2014singing}. 
To overcome this issue, some tried to consider the phase when separating the sources~\cite{kameoka2009complex, dubey2017does, le2019phasebook}\footnote{Using the full complex STFT number, instead of utilizing phaseless representations (either at the input or when applying the masks).}, or some others relied on a sinusoidal signal model at synthesis time~\cite{virtanen2000separation}.
However, in our work, we do not want to rely on any time-frequency transform or any signal model. Instead, we aim to directly approach the problem in the waveform domain.

As seen, many issues still exist around the idea of discarding the phase: are we missing crucial information when discarding it? When using the phase of the mixture at synthesis time, are we introducing artifacts that are limiting our model's performance? Or, since magnitude spectrograms (differently from power spectrograms) are not additive, which is the effect of relying on an incorrect signal model?

Our goal is to address these historical challenges via bypassing the problem. We want to investigate the feasibility of counting on an end-to-end model instead of relying on any signal model, any time-frequency transform, or filtering any signal. However, waveforms are high-dimensional and very variable. Thus, is music source separation possible in the waveform domain?
Recent literature shows that deep learning models operating on raw audio waveforms can achieve satisfactory results for several audio-based tasks~\cite{oord2017parallel,pons2018randomly,pons2017end}. 
And, among those, some are also recently starting to address the problem of music source separation directly in the waveform domain~\cite{stoller2018wave,grais2018raw}. 
Stoller et al.~\cite{stoller2018wave} proposed the Wave-U-Net (see Section 2.3 for more information), and Grais et al.~\cite{grais2018raw} proposed a multi-resolution\footnote{It is multi-resolution in the sense that they use several CNN filter lengths at every layer so that short- and long-term features can be efficiently learned/encoded. For further information, see Pons et al.~\cite{pons2018randomly}.} CNN auto-encoder for singing-voice source separation.
Unfortunately, though, these recent articles do not include any perceptual study comparing waveform-based models with spectrogram-based ones. One of our goals is to cover this literature gap to further understand which might be the impact of addressing music source separation in an end-to-end learning fashion. 
To this end, we set Wave-U-Net\footref{best} as one of our baselines and run a perceptual study to get a broader picture of how end-to-end learning models can perform.

As seen, the idea of approaching the music source separation task directly in the waveform domain has not been widely explored throughout the years, possibly due to the complexity of dealing with waveforms (which are unintuitive and high-dimensional). Consequently, during the last decades, music source separation in the waveform domain has been considered almost unattainable.
Our work aims to keep adding knowledge on top of this rather scarce literature, to convince the reader that music source separation is possible in the waveform domain.
To this end, in section~2 we first introduce a new end-to-end learning model: a Wavenet for music source separation. Later, we present two recent deep learning models that we set as baselines for our study: DeepConvSep~\cite{chandna2017monoaural} and Wave-U-Net~\cite{stoller2018wave}. 
In sections 3 and 4 we evaluate the above mentioned models, to conclude in section 5 that performing music source separation in the waveform domain is not only possible, but it can be a promising research direction. 
Hence, our main contributions can be summarized as follows:

\noindent \hspace{1mm} \textit{1)} We propose to use a Wavenet-based model for music source separation. Besides, we study the impact of several Wavenet hyper-parameters --- a result that might also be of relevance for other application areas where Wavenet has been used.

\noindent \hspace{1mm} \textit{2)} We perceptually benchmark several music source separation models, including our Wavenet-based model. This first perceptual study helps to further understand which might be the contribution of end-to-end models to the field of source separation.

\section{End-to-end source separation models}

We aim to discuss the feasibility of end-to-end learning models for monaural music source separation. To this end, we experiment with a new Wavenet-based model for music source separation we propose, and we compare it against two recent models: DeepConvSep~\cite{chandna2017monoaural}, a spectrogram-based deep learning model for multi-instrument separation; and Wave-U-Net~\cite{stoller2018wave}, a waveform-based model trained end-to-end for singing voice separation\footnote{\label{best}At the time of writing, DeepConvSep \& Wave-U-Net are the best \mbox{performing publicly available models for monaural music source separation.}}. We will compare the performance of these models perceptually and via assessing their {BSS Eval} scores \cite{evaltoolbox}. To allow a fair comparison, all discussed models are trained with MUSDB data down-sampled at 16kHz\footnote{DeepConvSep was trained by the original authors with DSD100 data~\cite{SiSEC16} at 44.1kHz. MUSDB is mostly conformed by DSD100.}.

\vspace{-2mm}
\subsection{A Wavenet-based model for source separation}

We utilize an adaptation of Wavenet \cite{van2016wavenet} that turns the original causal Wavenet (that is generative and slow), into a non-causal model (that is discriminative and parallelizable). This idea was originally proposed by Rethage et al.~\cite{rethage2017wavenet} for speech denoising, and we adapt it for monaural music source separation. Figure~1 shows an overall depiction of the model, where we can observe that every layer has residual and skip connections.
Before anything else, the waveform is linearly projected to \textit{k} channels by a 3x1 CNN-layer to comply with the feature map dimensions of each residual layer. Then, this projection is processed with several layers conformed by a dilated CNN passing through a tanh non-linearity controlled by a sigmoidal gate, see Figure~1~(\textit{Left}). The dilation factor in each layer increases in the range of 1, 2, ..., 256, 512. This ten layer pattern is repeated \textit{N} times (\textit{N} stacks).  
Later, two CNN layers (with \textit{k} filters, as well) adapt the resulting feature map dimensions to be the same as the residual and skip connections. A ReLU is applied after summing all skip connections and the final two 3x1 CNNs are not dilated~--- they have 2048 \& 256 filters, respectively, and are separated by a ReLU. The output layer linearly projects this feature map into as many channels as sources we aim to separate by using 1x1 filters. For multi-instrument source separation, our model has 3 outputs; and for singing voice separation, it has 1 single output. The remaining sources are computed via substracting the estimated sources from the mixture, see Figure 1 (\textit{Right}).
The main difference between the original Wavenet and the non-causal adaptation we use, is that some samples from the future can be used to predict the present one. 
As a result of removing the autoregressive causal nature of the original Wavenet, this fully convolutional model is able to predict a target field instead of one sample at a time~--- due to this parallelization, it is possible to run the model in real-time on a GPU \cite{rethage2017wavenet}.  
Another major difference with the original Wavenet is the output: we directly regress the waveform sources instead of sampling from a softmax output~\cite{rethage2017wavenet}. 
We minimize the mean absolute error (MAE) regression loss during training. ADAM optimizer is used with a learning rate of 0.001. We set the batch size to~10, and the model is trained until the validation error does not decrease for 16 epochs. The model with the lowest validation loss is selected. The code is accessible online.\footnote{\href{http://github.com/francesclluis/source-separation-wavenet}{https://github.com/francesclluis/source-separation-wavenet/}} 

\vspace{-2mm}
\subsection{DeepConvSep: a spectrogram-based model}

DeepConvSep \cite{chandna2017monoaural} is a state-of-the-art spectrogram-based model that is openly available\footnote{\href{https://github.com/MTG/DeepConvSep}{https://github.com/MTG/DeepConvSep/}}. Following the common practice: mixture signals (pre-processed as magnitude spectrograms) are fed to the model to estimate time-frequency soft masks for each source~\cite{stoter20182018,jansson2017singing,huang2014singing}. These masks are then used to filter the magnitude spectrogram of the mixture to estimate the magnitude spectrograms of the separated sources. Finally, these estimates, along with the phase of the mixture, are used to obtain the waveform signals corresponding to the separated sources.
DeepConvSep's architecture is based on a convolutional encoder-decoder. The encoder is conformed by a first CNN layer with 50 vertical filters aiming to capture timbral representations~\cite{pons2017timbre}, a second CNN layer with 30 horizontal filters modeling temporal cues~\cite{pons2017designing}, and a dense layer with 128 units acting as a bottleneck. The decoder contains two deconvolutional layers which up-sample the bottleneck feature maps up to have the same input size, which correspond to the estimated masks. The model learns via minimizing the mean squared error (MSE), together with several dissimilarity loss terms \cite{chandna2017monoaural,huang2014singing}. We utilize the original model released by the authors, as it is, which was trained with audio at 44kHz. To allow a fair comparison among models, we downsample its predictions to 16kHz~(which does not largely affect its performance, see Table~2).

\vspace{-2mm}
\subsection{Wave-U-Net: a waveform-based model}

Wave-U-Net \cite{stoller2018wave} is a state-of-the-art waveform-based model that is openly available\footnote{\href{https://github.com/f90/Wave-U-Net}{https://github.com/f90/Wave-U-Net}}. 
Wave-U-Net is a time-domain adaptation of the U-Net architecture for image segmentation~ \cite{ronneberger2015u}. It also consists in an encoder-decoder architecture. The encoder (12 layers) successively down-samples the feature maps, and the decoder (12 additional layers) up-samples the feature maps up to have the required output-length. A fundamental aspect of U-net architectures is that each decoder layer can access to the feature maps computed by the encoder (at the same level of hierarchy). To put an example: the penultimate layer, a decoder layer, can access the second layer's feature maps, an encoder layer, since they are concatenated. As a result of allowing the decoder to make use of the encoder feature maps, the output of the model is more detailed. These details come from the encoder-decoder connections, that convey the structure of the input to the output.
In order to allow a proper comparison among models, we re-train Wave-U-Net (following the best setup reported by the original authors: M3~\cite{stoller2018wave}) with MUSDB data at 16kHz, to fit the same train conditions as the Wavenet-based model.
We minimize the MSE loss during training. ADAM optimizer is used with a learning rate of 0.0001, and we set the batch size to 10. The model is trained until the validation error does not improve for 16 epochs. We select the model with the lowest validation loss.

\vspace{-1mm}
\section{Multi-instrument source separation}

The goal of this experiment is two-fold: \textit{i)} compare the proposed Wavenet-based model with DeepConvSep for the task of monaural multi-instrument source separation; and  \textit{ii)}~study several Wavenet hyper-parameter choices --- that are listed below:

\noindent \textbf{Wavenet: wider or deeper?}  
Provided that the GPU's memory is limited, this experiment explores the trade-off between how many filters each Wavenet layer has (the GPU's memory mostly stores learnable parameters with a wider Wavenet) and the receptive field length of the network (the GPU's memory mostly stores feature maps with a deeper Wavenet having a larger receptive field). Table~1 describes the setups we study.

\noindent \textbf{Which cost?} 
For our basic model we consider a single-term loss:
\(\mathcal{L}_{MAE} = \sum_{j\in J} \mid \hat{y}_{j}-y_{j} \mid\), where \(\hat{y}_{j}\) is the predicted source.
However, previous work successfully reduced interferences from other sources (SIR) via adding a dissimilarity loss term \cite{chandna2017monoaural,huang2014singing}:
\(\mathcal{L}_{d} = \sum_{j\in J} \sum_{i\in J}\mid{\hat{y}_{j}-y_{i\neq{j}}}\mid\),  with the resulting cost being: \(\mathcal{L}_{total} = \mathcal{L}_{MAE} - \alpha\cdot\mathcal{L}_{d}\). 
Small $\alpha$'s tend to perform well, and in our experiments we set $\alpha = 0.05$.
\vspace{-2mm}
\begin{table}[!h]
	\centering
	\caption{\textit{Description of the models we study. Wavenet-based ``\textit{k} filters" stand for the number of CNN filters in each residual connection, skip connection, and dilated convolutional~block.}}
	\vspace{-2mm}	
	\resizebox{\columnwidth}{!}{
		\begin{tabular}{c||c|c|c|c}
			\textit{Wavenet-based} & \textbf{ \textbf{\textit{k}}} & \textbf{\#} & \textbf{receptive} & \textbf{target} \\
			{\textit{N} stacks / layers} & \textbf{ filters} & \textbf{params} & \textbf{field} & \textbf{field} \\ \hline
			1 stack / 10 & 512 & $\approx$ 25.7M & 128 ms & 100 ms\\ 
			2 stacks / 20 & 256 & $\approx$ 13.6M & 256 ms& 100 ms\\ 
			3 stacks / 30 & 128 & $\approx$ 6.3M & 384 ms& 100 ms\\ 
			4 stacks / 40 & 64 & $\approx$ 3.3M & 512 ms& 100 ms\\ 
			5 stacks / 50 & 32 & $\approx$ 2.2M & 639 ms& 100 ms\\  \hline \hline
			\textit{DeepConvSep} & - & $\approx$ 314K & 290 ms& 290 ms \\ 
			\textit{Wave-U-Net} & - & $\approx$ 10.2M & 9.21 s & 1.02 s\\ 
	\end{tabular}}
\end{table}	

\vspace{-2mm}
\noindent Perceptual tests were conducted with 15 participants to get subjective feedback. 
Five songs were randomly chosen from 1' to 1'10'' to compose the perceptual test set.\footnote{\label{listen}\mbox{Listen: \href{http://jordipons.me/apps/end-to-end-music-source-separation/}{jordipons.me/apps/end-to-end-music-source-separation/}}} Participants were asked to ``\textit{give an overall quality score, taking into consideration both: sound quality of the target source and interferences from other sources}" for each of the estimated sources. The original mixture and the clean target source were presented as references. Participants provided a score between 1--5, with 1 being ``\textit{very intrusive interferences from other sources and degraded audio}", and 5 being ``\textit{unnoticeable interferences from other sources and not degraded audio}". Mean opinion score (MOS) is obtained by averaging the scores from all participants. 

\begin{table}[!h]
	\centering
	\vspace{-10mm}			
	\caption{\textit{Multi-instrument source separation median scores.}}
	\label{t:multi}
	\vspace{-3mm}
	\resizebox{\columnwidth}{!}{%
		\begin{tabular}{c c c c c c c}
			& \multicolumn{3}{c}{\textbf{\textit{Vocals}}} & \multicolumn{3}{|c}{\textbf{\textit{Drums}}}\\
			\multicolumn{1}{l}{{\textit{Wavenet-based}}} & \textbf{SDR} & \textbf{SIR} & \textbf{SAR} & \multicolumn{1}{|c}{\textbf{SDR}} & \textbf{SIR} & \textbf{SAR}\\ \hline
			\multicolumn{1}{l}{1 stack} & 0.35 & 3.94 & 4.38 & \multicolumn{1}{|c}{1.24} & 7.98 & 3.56\\
			\multicolumn{1}{l}{2 stacks} & 0.07 & 4.48 & 3.49 & \multicolumn{1}{|c}{ -0.09 } & 6.87 & 2.88\\
			\multicolumn{1}{l}{3 stacks} & \textbf{3.46} & 11.26 & 5.18 & \multicolumn{1}{|c}{\textbf{4.39}} & 13.37 & 5.08\\
			\multicolumn{1}{l}{4 stacks} & \textbf{3.35} & 11.25 & 5.24 & \multicolumn{1}{|c}{4.13} & 13.23 & 5.00\\
			\multicolumn{1}{l}{5 stacks} & 2.84 & 9.56 & 5.20 & \multicolumn{1}{|c}{\textbf{4.60}} & 12.66 & 6.08\\ \hdashline
			\multicolumn{1}{l}{4 stacks + $\mathcal{L}_{d}$} & 3.05 & 10.58 & 4.80 & \multicolumn{1}{|c}{4.09} & 12.85 & 5.31\\ \hline
			\multicolumn{1}{l}{\textit{{DeepConvSep 16kHz}}} & 2.38 & 4.45 & 8.39 & \multicolumn{1}{|c}{3.19} & 6.69 & 6.58\\ 
			\multicolumn{1}{l}{\textit{{DeepConvSep 44kHz}}} & 2.37 & 4.65 & 8.04 & \multicolumn{1}{|c}{3.14} & 6.73 & 6.55\\
			\midrule
			\midrule
			& \multicolumn{3}{c}{\textbf{\textit{Bass}}} & \multicolumn{3}{|c}{\textbf{\textit{Other}}}\\
			\multicolumn{1}{l}{{\textit{Wavenet-based}}} & \textbf{SDR} & \textbf{SIR} & \textbf{SAR} & \multicolumn{1}{|c}{\textbf{SDR}} & \textbf{SIR} & \textbf{SAR}\\ \hline
			\multicolumn{1}{l}{1 stack} & 0.35 & 4.54 & 4.70 & \multicolumn{1}{|c}{-2.70} & -1.37 & 6.75\\
			\multicolumn{1}{l}{2 stacks} & -0.55 & 0.87 & 7.80 & \multicolumn{1}{|c}{-2.05} & -0.97 & 8.96\\
			\multicolumn{1}{l}{3 stacks} & {2.24} & 6.36 & 5.94 & \multicolumn{1}{|c}{\textbf{0.54}} & 4.07 & 4.41\\
			\multicolumn{1}{l}{4 stacks} & \textbf{2.49} & 6.53 & 5.77 & \multicolumn{1}{|c}{\textbf{0.41}} & 3.83 & 4.47\\
			\multicolumn{1}{l}{5 stacks} & \textbf{2.48} & 6.70 & 6.27 & \multicolumn{1}{|c}{0.18} &  3.26 &  4.75\\ \hdashline
			\multicolumn{1}{l}{4 stacks + $\mathcal{L}_{d}$} & 2.23 & 5.66 & 6.37 & \multicolumn{1}{|c}{-0.19} & 4.37 & 3.24\\ \hline
			\multicolumn{1}{l}{\textit{{DeepConvSep 16kHz}}} & 0.27 & 1.92 & 7.46 & \multicolumn{1}{|c}{-2.02} & 1.74 & 2.50\\
			\multicolumn{1}{l}{\textit{{DeepConvSep 44kHz}}} & 0.17 & 1.98 & 7.06 & \multicolumn{1}{|c}{-2.13} & 1.84 & 2.33\\
	\end{tabular}}
    \vspace{-3mm}
\end{table}

\begin{table}[!h]
	\centering
	\caption{\textit{\mbox{Multi-instrument source separation perceptual scores.}}}	
	\vspace{-2mm}	
	\small {%
		\begin{tabular}{cccc}
			\textbf{MOS} & \textbf{\textit{Vocals}} & \textbf{\textit{Drums}} & \textbf{\textit{Bass}} \\ \hline
			\textit{Wavenet-based} & 2.4 $\pm$ 0.9 & 2.9 $\pm$ 1.1 & 2.4 $\pm$ 1.0 \\ 
			\textit{DeepConvSep 16kHz} & 2.3 $\pm$ 0.9 & 2.5 $\pm$ 0.7 & 1.8 $\pm$ 0.8 \\ 	
	\end{tabular}}
	\vspace{-6mm}
\end{table}

Table 2 shows the results of our experiments. Wide (but less deep) architectures fail at solving the task, only models having more than 3 stacks are able to perform competently. Two reasons may exist for that: wide models (having $>$~10M parameters) overfit the training set, and/or the small receptive field of wide models is not enough to solve the task. Further, we observe that the dissimilarity loss term $\mathcal{L}_{d}$ does not help improving the results.
Consequently, we choose the best performing model (4 stacks) for the perceptual test.
Table~3 presents the results of the perceptual test, showing that participants preferred the separations\footref{listen} done by the Wavenet-based model, particularly for drums \mbox{(t-test: p-value=0.018)} and bass \mbox{(t-test: p-value$<$$10^{-3}$)}. However, participants did not show any preference for the vocals' separations \mbox{(t-test: p-value=0.423).} This trend is consistent with BSS Eval scores, what shows that is possible to achieve good separations with end-to-end music source separation techniques.
Informal listening\footref{listen} also reveals that DeepConvSep is very conservative, possibly due to the mask-based approach used for filtering the spectrograms. Although Wavenet-based models seem to better remove the accompanying sources, they do it at the cost of introducing some artifacts that are noticeable when listening to the samples. 

\vspace{-1mm}
\section{Singing voice source separation}

The goal of this experiment is two-fold: \textit{i)} compare the proposed Wavenet-based model with Wave-U-Net for the task of monaural singing voice source separation; and  \textit{ii)} study several Wavenet hyper-parameter choices. Besides running \mbox{\textbf{Wavenet: wider or deeper?}} and \textbf{Which cost?} experiments for this setup, we extend our study with an extra experiment:

\noindent \textbf{Data-sampling strategies?} 
Our model had difficulties in producing continuous vocals. For that reason, we study training it using a higher proportion of the data containing singing voice (instead of vocal streams having silence). To study the contribution of this parameter, the percentage of forced fragments containing singing voice is set to~$[0, 25, 50, 75, 100]$ --- with 0\% meaning that segments are randomly selected, and 100\% meaning that our sampling strategy ensures that all examples contain singing voice.

\begin{table}[!t]
	\centering
	\vspace{-10.5mm}			
	\caption{\textit{\mbox{Singing voice source separation median scores.}}}
	\label{t:singing}	
	\vspace{-3mm}		
	\resizebox{\columnwidth}{!}{%
		\begin{tabular}{c c c c c c c}
			& \multicolumn{3}{c}{\textbf{\textit{Vocals}}} & \multicolumn{3}{|c}{\textbf{\textit{Accompaniment}}}\\
			\multicolumn{1}{l}{{\textit{Wavenet-based}}} & \textbf{SDR} & \textbf{SIR} & \textbf{SAR} & \multicolumn{1}{|c}{\textbf{SDR}} & \textbf{SIR} & \textbf{SAR}\\ \hline
			\multicolumn{1}{l}{1 stack} & 2.76 & 10.11 & 4.78 & \multicolumn{1}{|c}{9.73} & 12.73 & 13.77\\
			\multicolumn{1}{l}{2 stacks} & 3.05 & 11.13 & 4.50 & \multicolumn{1}{|c}{10.13} & 13.82 & 12.93\\
			\multicolumn{1}{l}{3 stacks} & 3.62 & 12.33 & 4.96 & \multicolumn{1}{|c}{10.41} & 13.97 & 13.53\\
			\multicolumn{1}{l}{4 stacks} & \textbf{3.67} & 12.14 & 5.24 & \multicolumn{1}{|c}{\textbf{10.64}} & 14.43 & 13.22\\
			\multicolumn{1}{l}{5 stacks} & 3.02 & 12.44 & 4.44 & \multicolumn{1}{|c}{10.42} & 13.89 & 13.30\\ \hdashline
			\multicolumn{1}{l}{4 stacks+$\mathcal{L}_{d}$} & 3.78 & 11.76 & 5.44 & \multicolumn{1}{|c}{10.90} & 14.26 & 13.84\\
			\multicolumn{1}{l}{4 stacks+$\mathcal{L}_{d}$+25\%} & 3.98 & 12.20 & 5.19 & \multicolumn{1}{|c}{10.75} & 14.21 & 13.70\\
			\multicolumn{1}{l}{4 stacks+$\mathcal{L}_{d}$+50\%} & \textbf{4.49} & 13.52 & 6.17 & \multicolumn{1}{|c}{\textbf{11.39}} & 16.37 & 13.49\\
			\multicolumn{1}{l}{4 stacks+$\mathcal{L}_{d}$+75\%} & 3.93 & 12.93 & 5.40 & \multicolumn{1}{|c}{11.14} & 16.18 & 13.37\\
			\multicolumn{1}{l}{4 stacks+$\mathcal{L}_{d}$+100\%} & 2.36 & 6.25 & 5.88 & \multicolumn{1}{|c}{10.44} & 16.73 & 12.15\\ \hline
			\multicolumn{1}{l}{\textit{{Wave-U-Net}}} & \textbf{4.60} & 14.30 & 5.54 & \multicolumn{1}{|c}{\textbf{11.87}} & 16.08 & 14.20\\
	\end{tabular}}
	\vspace{-3mm}
\end{table}
\begin{table}[!t]
	\centering
	\caption{\textit{\mbox{Singing voice source separation perceptual scores.}}}
	\vspace{-2mm}		
	\small {%
		\begin{tabular}{ccc}
			\textbf{MOS} & {\textit{Wavenet-based}} & \textit{Wave-U-Net} \\ \hline
			\textbf{\textit{Vocals}} & 3.0 $\pm$ 1.0 & 3.3 $\pm$ 0.85 \\ 
			\vspace{-9mm}			
	\end{tabular}}
	
\end{table}	

Table 4 shows the results of our experiments. Again, architectures having 3--4 stacks tend to perform better. However, differently from our previous experiment, the model having 1 stack performs reasonably. 
Further, we observe that the dissimilarity loss term $\mathcal{L}_{d}$ does help. And finally, note that carefully selecting the way we present our data to the model can make the difference. Our results greatly improve when 50\% of the training examples contain voice.
Consequently, we choose the best performing model (4~stacks+$\mathcal{L}_{d}$+50\%) for the perceptual test.
Table 5 presents the results of the perceptual test, showing that participants preferred Wave-U-Net separations\footref{listen} over Wavenet-based ones (t-test: \mbox{p-value}=0.049). This trend is consistent with BSS Eval scores, which denotes how powerful U-net architectures are for source separation~\cite{jansson2017singing,stoller2018wave}. That said, the remarkable performance of the proposed Wavenet-based model also indicates the potential of end-to-end music source separation models in general. Informal listening\footref{listen} also reveals that Wavenet-based models seem to better remove the accompanying sources. Although Wave-U-Net has difficulties in producing silences in parts having only accompaniment, its separations are smoother and have less artifacts --- that's why these separations are preferred by the listeners.
Finally, end-to-end models trained only for singing voice separation achieve much better results than their counterparts trained for multi-instrument separation \mbox{(compare Tables 4 and 5, against Tables 2 and 3).}

\vspace{-1mm}
\section{Discussion}

Throughout the years, end-to-end music source separation has been considered a hard research problem. Possibly because waveforms are variable and high-dimensional, the research community has focused on processing spectrograms instead of waveforms. 
However, with the recent advances of deep learning, music source separation starts to be possible in the waveform domain. 
As seen, although end-to-end music source separation methods have only started to be explored, the encouraging results we report denote the potential of this research direction --- that might, e.g., allow to bypass the inherent phase problems associated with some spectrogram-based methods, or to move beyond the current mask-based filtering paradigm.
To further show the viability of this research direction, we proposed a novel end-to-end source separation model based on Wavenet, that performs comparably to Wave-U-Net.
However, these two state-of-the-art waveform-based models perform $\approx$1.5dB (SDR) worse than the best spectrogram-based models that were published during the last SiSEC~(Signal Separation Evaluation Campaign~\cite{stoter20182018}).
Hence, although being possible and conceptually promising, end-to-end music source separation is still a challenging research topic.
Finally, as an additional way to validate the direction we explored, 
it is worth mentioning that the speech source separation community is also starting to propose end-to-end methods with some degree of success~\cite{venkataramani2017end,venkataramani2018end,luo2017tasnet}.

\section{Acknowledgements}

Work funded by the Maria de Maeztu Programme (MDM-2015-0502). We are grateful to NVidia for the donated GPUs.

\bibliographystyle{IEEEtran}

\bibliography{refs}

\end{document}